\documentclass[11pt]{amsart}
 
\theoremstyle{plain}
\newtheorem{thm}{Theorem}[section]

\theoremstyle{definition}
\newtheorem{remark}[thm]{Remark}

\numberwithin{equation}{section}

\title [A model for the measurement problem]{A model for measurements in Quantum Mechanics}
 \author{Tuyen Trung Truong}
        \date{\today}
        \address{School of Mathematics, Korea Institute for Advanced Study, Seoul 130-722, Republic of Korea}\email{truong@kias.re.kr}
 
\begin{document}

\maketitle

\begin{abstract}
Let $V=\mathbb{C}^N$, and $H$ (an observable) a Hermitian linear operator on $V$. Let $v_1,\ldots ,v_n$ be an orthonormal basis for $V$. Let $\mathcal{M}$ be a measurement apparatus prepared to measure a state of an observed system and collapses the state to  one of the $v_j$'s. Here we propose a simple model which explains the Born rule and is compatible with entanglement. 
\end{abstract}

Instead of its compatibility with experiments and many applications, the probability nature of Quantum Mechanics is still very much mysterious. In Quantum Mechanics (see, for example \cite{dirac, hughes}), an observable is a linear operator $H :V\rightarrow V$, where $V$ is a complex Hilbert space. The operator $H$ is assumed to be Hermitian, which means that $H=H^{\dagger}$ where $H^{\dagger}$ is the conjugate transpose of a complex matrix $H$. Moreover, it is assumed that the eigenvectors $\{v_i\}_{i\in I}$ of $H$ form a basis for $V$. An eigenstate of $H$ is an eigenvector of $H$. A state (or a ket)  is a vector $\xi\in V$. We view vectors $v$ and $c v$ as representing the same state, for any $c\not= 0$ a complex number. 

For simplicity, in the rest of this paper we will discuss only the case where $V$ is of finite dimensional. Therefore, we fix $V=\mathbb{C}^N$ with the usual inner product $<.,.>$ and $H:\mathbb{C}^N\rightarrow \mathbb{C}^N$ a Hermitian linear operator. We denote by $\lambda _1,\ldots ,\lambda _m$ the distinct eigenvalues of $H$, and by $V_j$ ($j=1,\ldots ,m$) the corresponding eigenspaces. We denote by $\mathcal{U}(n)$ the group of unitary linear operators, i.e. linear operators $U$ on $\mathbb{C}^N$ such that $U^{\dagger}U=UU^{\dagger}=Id$. Here $U^{\dagger}$ is the conjugate transpose of a complex matrix $U$.  

The most intriguing feature of Quantum Mechanics is the contrast between the smooth evolution of Schr\"odinger's evolution on one side, and the abruptly behavior under measurements on the other side. For the convenience of the readers, we will give a summary of these in the next. 

A state $v$ may involve in time, so we denote by $v(t)$ its dependence on time. Let $H(t)$ be the Hamilton of the system at time $t$, then Schrodinger's equation asserts that the state $v(t)$ will vary according to the rule
\begin{eqnarray*}
i\hbar\frac{d}{dt}v=H(t)v,
\end{eqnarray*}
here $\hbar$ is the Planck constant. In particular, if $H(t)=H$ is constant then the evolution of $v(t)$ is unitary and smooth
\begin{eqnarray*}
v(t)=e^{-\frac{i}{\hbar}Ht}\varphi ,
\end{eqnarray*}
where $\varphi =v(0)$.  

If $\xi\in V$ is an eigenstate of eigenvalue $\lambda$ of $H$, then the result of a measurement of $\xi$ will certainly give the value $\lambda$. If $\xi $ is not an eigenstate, then by the collapse of wave functions, a measurement will behave very wild: 

1) The result we obtain will be one of the eigenvalues $\lambda _1,\ldots ,\lambda _m$. Moreover, we will obtain the value $\lambda _j$ with probability proportional to $|<\xi ,P_j\xi >|^2$, where $P_j:\mathbb{C}^N\rightarrow V_j$ is the projection onto the eigenspace $\lambda _j$.  

2) After the measurement, $\xi$ will become one vector in $V_j$. 

\begin{remark} The above rules are those of the Copenhagen's interpretation. The experiments that satisfy these rules are called experiments of the first kind. There are however, many more experiments that do not satisfy these rules. 
\label{Remark1}\end{remark}

Now the contrast between the Schrodinger's evolution and the evolution under a measurement comes from the superposition principle in Quantum Mechanics. This principle says that if a system can be in states $v_1,\ldots ,v_k$, then it can also be in the state $c_1v_1+\ldots +c_kv_k$ for any complex numbers $c_1,\ldots ,c_k$. Therefore if we measure a state $\xi$ which is not an eigenstate, then before the measurement $\xi$ varies in a deterministic manner, but after the measurement then $\xi$ varies randomly. 

The contrast above, known as the measurement problem, has been extensively discussed since the beginning of Quantum Mechanics, with many interpretations (see for example \cite{hughes, dirac, penrose, wikipedia}). Here we will follow the well-known approach that the combined "observed system"+"measurement apparatus" should satisfy the Schrodinger's evolution. One contemporary representative of this interpretation is the quantum decoherence theory. This approach regards the combined "observed system" + "measurement apparatus" as entangled in and even after the measurement process, and thus why after the measurement the wave function collapses has not been explained. 

The purpose of the current paper is to give an interpretation of how after the measurement we may regard the observed system and the measurement apparatus to be independent. We  arrive equations of what happens in the measurement process, as viewed from the combined "observed system"+"measurement apparatus", or from the observed system alone. The measurement process is modeled as having $N$ gates, having energies attached. The observed system will give energy to each gate depending on how close it is to the gate, and will choose to enter the gate that have the largest total energy. The Born's rule follows naturally, and the model is compatible with entanglement.  

Here we give the mathematical details. Recall that $V=\mathbb{C}^N$, and $H$ (an observer) is a Hermitian linear operator on $V$. We let $v_1,\ldots ,v_n$ be an orthonormal basis for $V$. A measurement apparatus $\mathcal{M}$ is designed to measure states of observed systems and reduce the state to one of the $v_j$'s. The states of $\mathcal{M}$ are in another vector space $W$. A system $\mathcal{S}$, with state $\xi$, will be measured by $\mathcal{M}$. The time interval of each measurement is taken to be $[0,1]$.

{\bf Notations.} For a linear subspace $Z$ of a vector space, we let $P_Z$ be the projection onto $Z$. We will use $<,>$ for the usual inner product on either $V$, $W$ or $V\otimes W$. 

\subsection{Schrodinger's evolution of $\mathcal{S}+\mathcal{M}$} 
\label{SubsectionSchrodingerEquation}

The result in this subsection is well-known. The combined system $\mathcal{S}+\mathcal{M}$, may be regarded as closed. As so, they have a definite Hermitian Hamiltonian $\widehat{H}$, and satisfies the Schrodinger's equation. Therefore, a vector $z\in V\otimes W$ will evolute as $\widehat{U}(t)z$ where $$i\hbar \widehat{U}'(t)=\widehat{H}\widehat{U}(t),~\widehat{U}(0)=Id.$$ That is, $\widehat{U}(t)=e^{-i\widehat{H}t/\hbar}$. 

\subsection{Gates} 
\label{SubsectionGates}

We can imagine how the measurement process evolves as follows. First, the measurement apparatus $\mathcal{M}$ registers the state $\xi$. This $\xi$, a state in $V$, corresponds to the vector space $\xi \otimes W\subset V\otimes W$. If $\xi\otimes W$ is to end up at one of $v_j$'s at the end of the measurement process, then it must be brought to one of the vector spaces which are brought to $v_j\otimes W$ under the unitary operator $U(1)$. That is, $\xi \times W$ must go through one of the gates $$W_j:=\widehat{U}(1)^{-1}(v_j\times W),$$
and then proceeds under the unitary operator $\widehat{U}(t)$. 

We denote by $P_{W_j}$ the projections $P_{W_j}:V\otimes W\rightarrow W_j$. 

{\bf Remark.} The procedure how the observed system choose a gate to go into will be discussed later. 

\subsection{Energies of the gates}
\label{SubsectionEnergy}

For each gate $W_j$ there is associated a real number $\rho _j$ called the energy of the gate. These energies will be changed after each measurement has been made. The details will be discussed more later. 

\subsection{Closeness of the observed system and the gates}
\label{SubsectionCloseness}

Whether $\xi$ chooses the gate $W_j$ to enter depends on two  factors: the energy of the gate $W_j$ before the measurement and how close is  $\xi$ to the gate $W_j$. Here we characterize this closeness as follows. Using that $\widehat{U}(t)$ is unitary, we define the closeness by how close $\widehat{U}(1)\xi \otimes W$ is to $v_j\otimes W$. The latter is proportional to the length of the trace of the projection from $\widehat{U}(1)\xi \otimes W$ to $v_j\otimes W$. The latter vector is
\begin{eqnarray*}
Tr_W(P_{v_j\otimes W}\widehat{U}(1))\xi =\sum _{i=1}^m<P_{v_j\otimes W}(\widehat{U}(1)\xi \otimes w_i),w_i>.
\end{eqnarray*}
Here $w_1,\ldots ,w_m$ are an orthonormal basis for $W$.

Therefore, the closeness of $\xi$ to the gate $W_j$ is proportional to 
\begin{eqnarray*}
|Tr_W(P_{v_j\otimes W}\widehat{U}(1))\xi |^2=|\sum _{i=1}^m<P_{v_j\otimes W}(\widehat{U}(1)\xi \otimes w_i),w_i>|^2.
\end{eqnarray*}

\subsection{Equations of the measurement process}
\label{SubsectionEquations}

From the point of view of the combined system $\mathcal{S}+\mathcal{M}$, when the result of a measurement is $v_j$, then  the value $v_j\otimes W$ is definite. Hence to characterize the measurement process, we need to go backward from $v_j\otimes W$ 
\begin{eqnarray*}
\widehat{U}_j(t):=\widehat{U}(t)\widehat{U}(1)^{-1}P_{v_j\otimes W}\widehat{U}(1). 
\end{eqnarray*}
Here, for a linear subspace $Z\subset V\otimes W$, $P_Z$ denotes the projection onto $Z$. We note that $\widehat{U}_j(t)$ also satisfies the Schrodinger's equation 
\begin{eqnarray*}
i\hbar \widehat{U}_j'(t)=\widehat{H}\widehat{U}_j(t).
\end{eqnarray*}
Moreover, since $v_1,\ldots ,v_N$ are an orthonormal basis for $V$, it follows that 
\begin{eqnarray*}
\sum _{j=1}^N\widehat{U}_j(t)=\widehat{U}(t).
\end{eqnarray*} 
 
From the point of view of the observed system $\mathcal{S}$ alone, the evolution is the trace of the $\widehat{U}_j(t)$'s, hence is characterized by
  $$U _j(t)\xi =Tr_W(\widehat{U}(t)\widehat{U}(1)^{-1}P_{v_j\otimes W}\widehat{U}(1))\xi =\sum _{i=1}^{m}<\widehat{U}(t)\widehat{U}(1)^{-1}P_{v_j\otimes W}(\widehat{U}(1)\xi \otimes w_i),w_i>.$$
Here, $w_1,\ldots ,w_m$ are an orthonormal basis for $W$.

\subsection{The measurement process} Here we describe in detail how a gate is chosen. 
\label{SubsectionProcess}

We recall that to each gate $W_j=\widehat{U}(1)^{-1}(v_j\otimes W)$, there associates a real number $\rho _j$ called its energy. We give the following complete ordering on the pairs $(W_j,\rho _j)$ ($j=1,\ldots ,N$). We define $(W_j,\rho _j)>(W_k,\rho _k)$ if: Either $\rho _j>\rho _k$, or $\rho _j=\rho _k$ and $j>k$.

When $\mathcal{M}$ measures an $\mathcal{S}$ with state $\xi$, then the following happens: 

1) The gates for which $Tr_W(P_{v_j\otimes W}\widehat{U}(1))\xi =0$ are disregarded.

2) For every $j=1,\ldots ,N$, $\rho _j$ is changed to $\rho _j+|Tr_W(P_{v_j\otimes W}\widehat{U}(1))\xi |^2$.

3) $\mathcal{S}$ will go into the gate $W_{j_0}$, where $(W_{j_0},\rho _{j_0})$ is the maximum element, under the ordering defined in the previous paragraph, among those with $|Tr_W(P_{v_j\otimes W}\widehat{U}(1))\xi |^2>0$. Then $\rho _{j_0} $ is changed to $\rho _{j_0}-1$.

Here we normalize $\xi$ so that 
\begin{eqnarray*}
\sum _{j=1}^N|Tr_W(P_{v_j\otimes W}\widehat{U}(1))\xi |^2=1.
\end{eqnarray*}

{\bf Remarks.} Here we give some remarks on the complete ordering on the pairs $(W_j,\rho _j)$. If there are two gates $W_{j_1}$ and $W_{j_2}$ such that $\rho _{j_1}=\rho _{j_2}$, then we need to choose an ordering of the gates $W_{j_1}$ and $W_{j_2}$ in order to choose one gate for the observed system to go through. This seems a bit unnatural. However, this may be avoided if the initial energies $\rho _j(0)$ of the gates, before any measurement was made at all, were so that  
\begin{eqnarray*}
\rho _j(0)+m_j+n_j|Tr_W(P_{v_j\otimes W}\widehat{U}(1))\xi |^2\not= \rho _k(0)+m_k+n_k|Tr_W(P_{v_k\otimes W}\widehat{U}(1))\xi |^2,
\end{eqnarray*}
for all $j\not= k$ and here $m_j,n_j,m_k,n_k$ are arbitrary integers. This condition is satisfied generically. There is one other way to avoid having to order the gates $W_{j_1}$ and $W_{j_2}$, by allowing the energies $\rho _j$ to be randomly changed every once in a while. See Subsection \ref{SubsectionRandomness} for more on this. 

\subsection{Born's rule} 
\label{SubsectionBorn}

Here we give a proof that the measurement process described in the above paragraph satisfies the Born's rule of the Copenhagen's interpretation. 

\begin{proof}
By the normalized condition 
\begin{eqnarray*}
\sum _{j=1}^N|Tr_W(P_{v_j\otimes W}\widehat{U}(1))\xi |^2=1,
\end{eqnarray*}
it follows that 
\begin{eqnarray*}
\sum _{j=1}^N\rho _j=C
\end{eqnarray*}
is unchanged, no matter how many measurements have been made.
Now assume that a measurement is made in the time interval $[0,1]$. We let $\rho _j$ be the value of the capacity of the gate $W_j$ at time $0$, and let $\rho _j'$ be the value of the capacity of the gate $W_j$ at time $1$. Let $\rho "_j=\rho _j+|Tr_W(P_{v_j\otimes W}\widehat{U}(1))\xi |^2$ be the values of the capacities after Step 2 of the measurement process. Then 
\begin{eqnarray*}
\sum _{j=1}^N\rho "_j=1+\sum _{j=1}^N\rho _j=C+1.
\end{eqnarray*}
Hence if $\rho _j<-|C|-2$, there will be one $\rho _k$ with $\rho "_k>0>\rho "_{j}$. It follows that the gate $W_j$ will not be chosen, and hence after the measurement $\rho '_j=\rho "_j\geq \rho _j$. 

It follows from this that there is a constant $B$ such that we always have
\begin{eqnarray*}
\rho _j>B
\end{eqnarray*}
for all $j$. In fact, before the measurement, assume that for all $j=1,\ldots ,N$ we have $\rho _j>B$, for some $B<-|C|-3$. Then after the measurement, all $\rho _j$ with $B<\rho _j<-|C|-2$ will have $\rho '_j\geq \rho _j\geq B$. There is only one $j_0$ with $\rho '_{j_0}\leq \rho _{j_0}$, but for that one then $\rho '_{j_0}\geq \rho _j-1\geq -|C|-2-1>B$. 

We also have an upper bound for $\rho _j$. In fact, for any $j=1,\ldots ,N$ we always have
\begin{eqnarray*}
\rho _j=\sum _{k=1}^N\rho _k-\sum _{k\not= j}\rho _k=C-\sum _{k\not= j}\rho _k<C-(N-1)B.
\end{eqnarray*}
 
Now let $n$ be a large integer, and let $n_j$ be the number of times that in $n$ measurements, the gate $W_j$ has been chosen $n_j$ times. Then after $n$ measurements we have
\begin{eqnarray*}
B<\rho _j=n|Tr_W(P_{v_j\otimes W}\widehat{U}(1))\xi |^2-n_j<C-(N-1)B.
\end{eqnarray*}
Therefore  
\begin{eqnarray*}
\lim _{n\rightarrow\infty}\frac{n_j}{n}=  |Tr_W(P_{v_j\otimes W}\widehat{U}(1))\xi |^2
\end{eqnarray*}
for all $j=1,\ldots ,N$. 
\end{proof}

\subsection{Independence of $\mathcal{S}$ and $\mathcal{M}$ after the measurement}
\label{SubsectionIndependence}

It is reasonable to regard that if at the end of a measurement, the Hamiltonian of the system $\mathcal{S}$ returns to $H$, then $\mathcal{S}$ is independent from $\mathcal{M}$. From the definitions of ${U}_j(t)$ we find that 
\begin{eqnarray*}
i\hbar U_j'(t)\xi =Tr_W(\widehat{H}\widehat{U}(t)\widehat{U}(1)^{-1}P_{v_j\otimes W}\widehat{U}(1))\xi .
\end{eqnarray*}
Hence, at $t=1$ 
\begin{eqnarray*}
i\hbar U_j'(t)\xi =Tr_W(\widehat{H}P_{v_j\otimes W}\widehat{U}(1))\xi .
\end{eqnarray*}
Hence, if 
\begin{eqnarray*}
Tr_W(\widehat{H}P_{v_j\otimes W}\widehat{U}(1))= HTr_W(P_{v_j\otimes W}\widehat{U}(1)), 
\end{eqnarray*}
for all $j=1,\ldots , N$, then we may regard $\mathcal{S}$ and $\mathcal{M}$ as independent after the measurement.   

\subsection{Quantum entanglement} 
\label{SubsectionEntanglement}
Here we show how the model can be extended to be compatible with quantum entanglement.

We recall briefly what is quantum entanglement. Here we have two observed systems $\xi _1\in V_1$ and $\xi _2\in V_2$ which are entangled in such a way that their states can not be written as $\xi _1\otimes \xi _2\in V_1\otimes V_2$, but rather by a general element $\xi \in V_1\otimes V_2$.  

The intriguing feature of quantum entanglement is that whenever $\xi _1$ is measured and collapses to an eigenstate $v_1$, then $\xi _2$ immediately becomes the eigenstate 
\begin{eqnarray*}
v_2=<v_1,\xi >.
\end{eqnarray*}
 
Here we describe the extended model. When a measurement apparatus $\mathcal{M}$ measures $\xi _1$, it somehow knows that $\xi _1$ is one part of an entangled pair. Hence, the end results are not $v_1,\ldots ,v_n$ any more, but are vectors in $v_1\otimes V_2,\ldots ,v_n\otimes V_2$. Then the corresponding gates will be changed to $v_1\otimes V_2\otimes W,\ldots ,v_n\otimes V_2\otimes W$. The energies on $v_j\otimes V_2\otimes  W$ will be the same as the energies on $v_j\otimes W$, which are from the original $\mathcal{M}$. Then we proceed the process as in the case considered before, where there is no entanglement (see Subsection \ref{SubsectionProcess}). 

\subsection{Randomness} 
\label{SubsectionRandomness}

From Subsection \ref{SubsectionProcess}, we see that from the viewpoint of the measurement apparatus $\mathcal{M}$, the measurement process is predictable. Having known the energies of the gates before a measurement is made, we will know the result of the measurement even before making it. However, from the viewpoint of an observer, the measurement process looks very much random. Here we offer some reasons for why an observer can not know the energies of the gates before hand: 

- One reason is that the measurement apparatus may make more measurements than an observer may know.

- One other reason is that occasionally, because of some unknown factors, the energies of the gates will be changed randomly. The observer is not aware of these random changes.  

\subsection{Further comments}

We may take the view that the energies are associated either to the measurement apparatus or to observed systems. However, it seems that these energies are more natural associated to the measurement apparatus than to observed systems. The reason is that before any measurement is made, the observed system has no preference on what states it will collapse to. Therefore, if it was to keep information on a certain orthonormal basis $v_1,\ldots ,v_N$, it must do so for all such orthonormal bases, and this will be somehow burdensome. Therefore, it is better that the information on the energies are kept within the measurement apparatus. 

The above model may be interpreted as follows. When an observed system is measured, then it provides a total $1$ unit of energy to the gates corresponding with how close it is to the gates. Thus, while the observed system goes through only one gate, it leaves imprints on all gates. This energy helps to make the observed system and the  measurement apparatus becoming entangled initially.  When $\mathcal{S}$ goes through the gate $W_{j_0}$, then to keep $\mathcal{S}$ stay inside and then to send out the result at the end, the gate $W_{j_0}$ must spend energy.  This is why $W_{j_0}$ should have the maximum energy, and why after the measurement then the energy of $W_{j_0}$ is decreased by $1$ unit. We see that energy is preserved in this measurement process. 

{\bf Generalizations.} The results above extend without difficulties to the case where $\mathcal{M}$ is designed to collapse the states to one of pairwise orthogonal linear subspaces $V_1,\ldots ,V_l$ of $V$.


\begin{thebibliography}{TTT}

\bibitem{hughes} R. I. G. Hughes, {\em The structure and interpretation of Quantum Mechanics,} Harvard University Press, 7th edition, 2003.

\bibitem{dirac} P. A. M. Dirac, {\em Quantum Mechanics,} Oxford University Press, 4th edition, 1958.

\bibitem{penrose} R. Penrose, {\em The road to reality,} Vintage Books, 2004.

\bibitem{wikipedia} The Wikipedia's pages on "Wave function collapse" and "Quantum decoherence". 
\end{thebibliography}
\end{document}